\documentclass[aps,prl,twocolumn,superscriptaddress,showpacs]{revtex4}
\usepackage{graphicx}
\usepackage{amssymb}
\begin{document}
\bibliographystyle{apsrev}

\title{Stiffness of Contacts Between Rough Surfaces}
\author{Sreekanth Akarapu, Tristan Sharp and Mark O. Robbins}
\affiliation{Department of Physics and Astronomy, Johns Hopkins University,
3400 N Charles St, Baltimore, MD 21218, USA}

\date{\today}

\begin{abstract}
The effect of self-affine roughness on solid contact is
examined with molecular dynamics and continuum calculations.
The contact area and normal and lateral stiffnesses rise linearly
with the applied load, and the load rises exponentially with
decreasing separation between surfaces.
Results for a wide range of roughnesses, system sizes and Poisson
ratios can be collapsed using Persson's contact theory for
continuous elastic media.
The atomic scale response at the interface between solids has little
affect on the area or normal stiffness, but can greatly reduce the
lateral stiffness.
The scaling of this effect with system size and roughness is discussed.

\end{abstract}


\maketitle

The presence of roughness on a wide range of length scales has profound
effects on contact and friction between experimental surfaces.
Under a broad range of conditions
\cite{bowden86,greenwood66,persson01,hyun04,pei05,campana07epl},
the area of intimate contact
between rough surfaces $A_c$ is orders of magnitude smaller
than the apparent surface area
$A_0$.
As discussed below, this provides the most common explanation for Amontons'
laws that friction is proportional to load and independent of $A_0$.
Because $A_c$ is small, the interfacial region is very compliant.
In a range of applications the interfacial compliance can
significantly reduce the stiffness of macroscopic joints
formed by holding two components together under pressure \cite{bowden86,bellow70}.

In this paper, we examine the effect of surface roughness on the
normal and lateral stiffness of contacts between elastic solids
using molecular dynamics (MD) and continuum calculations.
The results provide a numerical test of recent continuum theories
\cite{persson07,persson08c} and their applicability to real solids.
The contact area and normal stiffness approach continuum predictions
rapidly as system size increases.
Continuum theory also captures the internal deformations in the solid
under tangential load, but the total lateral stiffness may be greatly
reduced by atomic scale displacements between contacting atoms
on the opposing surfaces.
This makes it a sensitive probe of the forces underlying friction and
may help to explain unexpectedly small experimental results
\cite{experiments}.

The topography of many surfaces can be described as a self-affine
fractal \cite{archard57,greenwood66}.
Over a wide range of lengths, the root mean squared (rms)
change in height $dh$ over
a lateral distance $\ell$ scales as a power law: $dh \sim \ell^H$,
where the roughness or Hurst exponent $H$ is typically between 0.5 and 0.9.
Greenwood and Williamson (GW) considered the peaks of rough landscapes as
independent asperities and found that $A_c$ rose
linearly with normal load $F_N$ for nonadhesive surfaces \cite{greenwood66}.
This explains Amontons's laws if there 
is a constant shear stress at the interface.
A linear scaling of area with load is also obtained from Persson's
scaling theory, which includes elastic coupling between contacts
approximately \cite{persson00,muser08prl}.

Dimensional analysis implies that the linear relation between load
and area must have the form
\begin{equation}
A_c E'=\kappa F_N / \sqrt{| \nabla h|^2}  .
\label{kappa}
\end{equation}
where a modulus like the contact modulus $E'$ is the only dimensional quantity
characterizing the elastic response, and the rms slope the only
dimensionless quantity characterizing the roughness.
Numerical solutions of the continuum
equations \cite{hyun04,campana07epl}
show that
$\kappa$ is near 2.
Results for different $H$ and Poisson ratio $\nu$
lie between the analytic predictions of GW,
$\sqrt{2\pi}\sim2.5$, and Persson, $\sqrt{8/\pi}\sim1.6$.
One advantage of Persson's model is that, as in numerical results,
$A_c/F_N$ is constant
over a much larger range of loads than GW \cite{carbone08}.
Another is that it captures \cite{persson08c} the power law
scaling of correlations in contact and stress that was found in
numerical studies \cite{hyun07,campana08}.

The normal stiffness is related to the change in average surface separation
$u$ with load.
Experiments \cite{benz06,lorenz09} and
calculations \cite{pei05,yang08,persson07} show an exponential rise in load
with decreasing $u$,
$F_N =c A_0 E' \exp[-u/\gamma h_{\rm rms}]$,
where $h_{\rm rms}$ is the root mean squared (rms)
variation in surface height and $\gamma$ a constant of order 1.
Differentiating leads to an expression for the normal interfacial stiffness:
\begin{equation}
k^I_N=-dF_N/du = F_N/\gamma h_{\rm rms}.
\label{normalstiff}
\end{equation}
For self-affine surfaces this interfacial stiffness decreases as
$h_{\rm rms}^{-1} \sim L^{-H}$
with increasing system size $L$.
Our simulations test this scaling and show that $\gamma$ is nearly constant.
They also examine the connection between this normal stiffness and the
transverse stiffness $k^I_T$ at forces lower than the static friction 
\cite{mindlin49}.

We consider nonadhesive contact of a rigid rough solid and a flat elastic
substrate.
This can be mapped to contact of two rough, elastic solids in
continuum theories \cite{greenwood66,persson00}.
The mapping is only approximate for atomic systems \cite{luan05,luan09},
but working with one rigid solid reduces the parameter space in this
initial study.
Rigid surface atoms are placed on the sites of a simple
cubic lattice with spacing $d'$, and only atoms in the outer layer
are close enough to interact with the substrate.
They are displaced vertically to coincide with a self-affine fractal surface of the desired $H$.
Surfaces with roughness on wavelengths from $l_{\rm min}$ to $l_{\rm max}$
were generated as in Ref. \cite{hyun07}.
The rms slope $\sqrt{|\nabla h|^2} = 0.1$ for the results shown.
Consistent results were obtained for slopes from 0.05 to 0.15.
Slopes of 0.2 or greater led to plastic deformation in the substrate.
Continuum calculations also show plasticity for large slopes \cite{pei05}. 

The elastic substrate is an fcc crystal with
nearest-neighbor spacing $d$ and a (100) surface.
Surface atoms form a square lattice that is rotated by $45^\circ$
relative to that of the rigid surface.
This rotation and the choice of $d/d'$ prevent commensurate locking
of the surfaces \cite{hirano93,muser01prl}.
Substrate atoms separated by $r$
interact with a Lennard-Jones (LJ) potential:
$U_{LJ} = 4\epsilon [(\sigma/r)^{12}-(\sigma/r)^{6}]$,
where $\epsilon$ and $\sigma$ are the bonding energy and diameter.
To speed calculations, the potential and force are interpolated
smoothly to zero at $r_c=1.8\sigma$
\cite{lammps}.
Since we compare to continuum theories with no interfacial adhesion,
purely repulsive interactions are used between substrate and rigid atoms.
An LJ potential with length $\sigma '$
is truncated at the energy minimum, $2^{1/6}\sigma '$.
Unless noted
$\sigma'=0.5\sigma$ and $d'=0.3d$ to minimize interfacial friction and speed
the approach to continuum theory \cite{campana07epl}.

Substrate atoms are arranged to form a cube of side $L$.
Periodic boundary conditions are applied in the plane of the surface and
bottom atoms are held fixed.
The elastic constants $c_{33} = 70.2\epsilon/\sigma^3$ and
$c_{44} =41.8\epsilon/\sigma^3$ were measured by displacing
the top surface.
Cubic crystals do not exhibit the isotropic elasticity
assumed in continuum theory.
To test for any effect from anisotropy, we also performed Greens function
MD (GFMD) \cite{campana06}
for isotropic continua with $\nu=0$ and 0.35.
Since thermal fluctuations are ignored in continuum theory,
we consider the limit of zero temperature $T$.
The energy is minimized for given external forces or displacements.
The fractional contact area $A_c/A_0$ is obtained from the fraction
of surface atoms that interact with the rigid surface \cite{luan09}.

In all cases studied, $A_c$ rises linearly with $F_N$.
Moreover, 
the value of $\kappa$ approaches previous continuum results
as system size increases.
The decrease in $\kappa$ with increasing $L$ is similar to that
found by Campana and M\"user in their GFMD
calculations \cite{campana07epl}.
We found $\kappa$ converged more rapidly
when their atomistic Greens function
was replaced by an ideal elastic Greens function
and we use this in the GFMD results below.
As in previous 2D atomistic calculations \cite{luan09},
the stress and contact correlation functions
from our LJ calculations
exhibited the power law scaling $q^{-(1+H)}$ found in GFMD calculations
and Persson's theory \cite{campana08,persson08c}.

\begin{figure}[tb]
\centering
\includegraphics[width=2.5in]{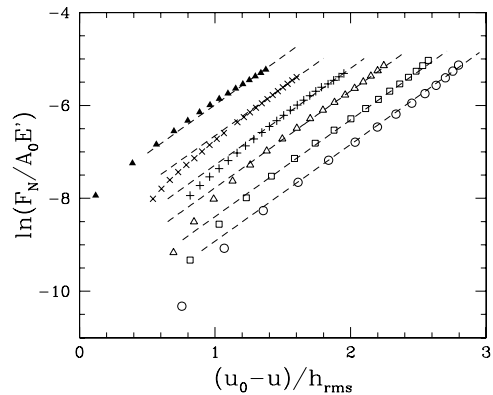}
\caption{
Logarithm of load as a function of $(u_0-u)/h_{\rm rms}$,
and linear fits corresponding to $\gamma=0.48$.
The separation at first contact, $u_0$, is shifted slightly
to prevent overlap.
Atomistic results are for $H=0.5$ with $L=378.4 \sigma$ (circles),
$189.2 \sigma$
(squares), and $94.6 \sigma$ (triangles) and for $H=0.8$ with
$L=189.2 \sigma$ (crosses) and $94.6\sigma$ (pluses).
Filled triangles are for a GFMD simulation with $L=128d$
and $\nu=0$.
}
\label{fig:spacing}
\end{figure}

Figure \ref{fig:spacing} shows the variation of $F_N$ with
interfacial separation $u$ for several $L$ and $H=0.5$ and $0.8$.
In all cases,
$F_N$ rises exponentially over a range of loads
that corresponds to fractional contact areas between 1 and 10\%.
Statistics are too poor at lower areas and nonlinear corrections to Eq. \ref{kappa} are seen at larger areas \cite{hyun04}.
The linear fits to all results have the same slope, corresponding to
$\gamma=0.48$ and best fit values for all $H$ and $L$ 
studied differ by less than 10\% from this value.
GFMD results were at the higher end of this range and showed no change as
$\nu$ increased from 0 to 0.35.
Earlier continuum calculations \cite{pei05}, elastic atomic
calculations \cite{yang08} and experiments \cite{lorenz09} were consistent
with $\gamma \approx 0.4$.
This represents a compelling success of Persson's approach, and
raises the question of whether $\gamma$ may have a unique value
in the thermodynamic, isotropic limit.

The normal stiffness from Eq. \ref{normalstiff}
includes a component from the increase in contact
area with load as well as the change in force at fixed area.
There is also a compliance associated with changes in the
separation between contacting rigid and substrate atoms that is
generally neglected in continuum theory.
To isolate the stiffness associated with deformation
within the substrate at fixed contact area $k^{Is}$,
we applied constraints directly to the atoms that contacted at a
given load.
The normal and lateral stiffness were then obtained by displacing
these contacting atoms
in the normal or lateral direction and measuring the change in force.
The contribution from the bulk response was subtracted so that
the stiffness reflects the change in surface separation $u$ or
lateral surface translation $u_T$.
This approach is straightforward to implement in experiments and
was found to be consistent with direct averaging of atomic separations.

\begin{figure}[tb]
\centering
\includegraphics[width=2.5in]{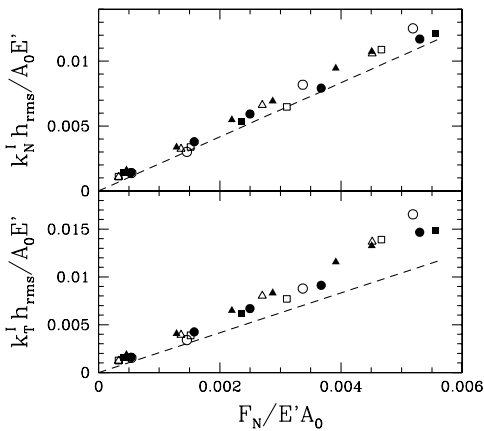}
\caption{
The scaled (a) normal stiffness and (b) tangential stiffness as
a function of $W/A_0E*$.
Results are for $H=0.5$ (open symbols) and $H=0.8$ (filled symbols)
with $L=189.2\sigma$ (circles), $L=94.6\sigma$ (squares) or
$L=47.3\sigma$ (triangles).
Dashed lines have slope $1/\gamma$ with $\gamma=0.48$.
}
\label{fig:stiffload}
\end{figure}

Figure \ref{fig:stiffload}(a) shows the scaled normal interfacial stiffness
$k^{Is}_N h_{rms}/A_0 E'$ as a function of the dimensionless load $F_N/A_0E'$
used to find the contacting atoms.
Once again, results for all systems show the same behavior,
and the stiffness rises linearly with load as predicted by
Eq. \ref{normalstiff}.
The points lie slightly above the dashed line corresponding to
$\gamma=0.48$ due to small deviations from the analytic form of Eq. \ref{normalstiff}.
One might expect $k^{Is}_N$ to be substantially less than the total
stiffness because it does not include the stiffness from increases
in contact area.
However, the two stiffnesses are nearly the same because newly contacting
regions carry the smallest forces.

The incremental response of an ideal elastic solid does not depend on any
pre-existing deformation.
This implies that we should obtain the same stiffness by displacing
the same set of atoms on the initial undeformed surface.
Direct evaluation of the stiffness in this way gave slightly lower stiffnesses
than Fig. \ref{fig:stiffload},
with the difference increasing from the numerical uncertainty
to about 15\% with increasing $F_N$.
This provides an estimate of the contribution that anharmonic
effects may make to the stiffness of real materials at the
rms slope used here.

The above results imply that the stiffness of elastic solids at
fixed contact area is uniquely determined by the distribution of contacting
points.
This conclusion may seem at odds with Eq. \ref{normalstiff},
since the contact area has no independent connection to load or surface
roughness.
The resolution is that variations in load and roughness cancel.
If the response is linear,
one can scale
$h_{\rm rms}$ and $F_N$ by the same factor and the contact area
will be unchanged.
Indeed one can combine Eqs. \ref{kappa}
and \ref{normalstiff} to eliminate $F_N$:
\begin{equation}
k_N^* \equiv \frac{k^I_N }{A_0 E'}\frac{h_{\rm rms}}{\sqrt{| \nabla h|^2}} 
 = \frac{1}{\kappa \gamma} \frac{A_c}{A_0} .
\label{eq:area}
\end{equation}
For a self-affine surface, the ratio
$h_{\rm rms}/\sqrt{|\nabla h|^2} \propto (l_{\rm max}/l_{\rm min})^H l_{\rm min}$
depends only on the small and large scale cutoffs in roughness.

\begin{figure}[b]
\centering
\includegraphics[width=2.4in]{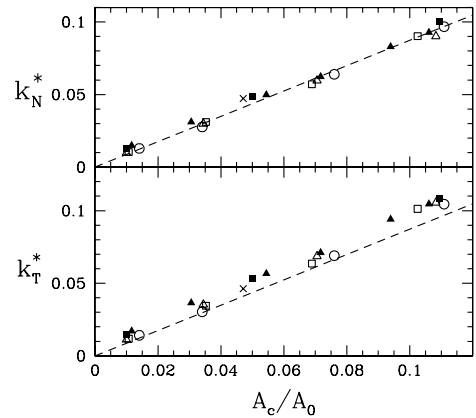}
\caption{
The scaled (a) normal stiffness and (b) tangential stiffness as
a function of $A_c/A_0$.
Results are for $H=0.5$ (open symbols) and $H=0.8$ (filled symbols)
with $L=189.2\sigma$ (circles), $L=94.6\sigma$ (squares) or
$L=47.3\sigma$ (triangles).
Crosses show GFMD results for $\nu=0$.
Dashed lines have slope 0.87.
}
\label{fig:stiffarea}
\end{figure}

Figure \ref{fig:stiffarea}(a) shows the scaled stiffness $k_N^*$
vs. area.
The results were obtained by displacing atoms from their positions on
the initial flat surface to eliminate anharmonic effects.
Results for all systems collapse onto a common straight line,
providing clear evidence for the direct connection between stiffness
and contacting area.
The slope is near unity as expected from the separate values of $\kappa$
and $\gamma$.

All of our atomic simulations show $k^{Is}_T/k^{Is}_N >1$.
This is surprising given that Mindlin \cite{mindlin49} and
recent work \cite{perssonpriv} predict
$k^I_T/k^I_N = 2(1-\nu)/(2-\nu)<1$.
However, this result is for isotropic systems.
One measure of the anisotropy of the LJ crystal is that the ratio
$c_{44}/E' \approx 0.57$, while it is $(1-\nu)/2 < 1/2$ for an
isotropic solid.
This is also consistent with shear stresses having a higher relative
stiffness than expected.
In general, the total elastic energy
stored in the interface is
$\Sigma_q \vec{f}(\vec{-q}) G(\vec{q}) \vec{f}(\vec{q})/2$ where
$G$ is the Greens function matrix relating displacements to forces $f$
\cite{campana06}.
The ratio of stiffnesses can be obtained by averaging the diagonal components
of $q G(\vec{q})$
corresponding to normal and tangential displacements over $\hat q$ and assuming
the same power spectrum describes the respective forces.
This ratio agrees with Mindlin's result for isotropic systems, but will
vary with crystal anisotropy.

As noted above, $k^{Is}_N$ and $k^{Is}_T$ only consider the stiffness
associated with deformation inside the substrate at constant area.
There is an additional stiffness $k^{Ia}$ associated with relative
motion at the interface
between contacting substrate atoms and the rigid surface.
Simulations of single asperity contacts show that this atomic scale
motion has very different effects on the total normal and tangential
contact stiffness \cite{luan05}.
The stiffness $k^{Ia}_N$ resisting normal displacements is large, because contacting
atoms are in the steep repulsive part of the LJ potential.
The compliance associated with these bonds has little effect on the
total stiffness of single asperities or the multiasperity contacts studied
here.
As shown above, the full normal stiffness including this compliance
and changes in area (Fig. \ref{fig:spacing})
is consistent with the stiffness from compression
of the elastic substrate alone (Fig. \ref{fig:stiffload}).

In contrast, the tangential stiffness of single asperity contacts
can be an order of magnitude or more smaller than continuum
predictions \cite{luan05}.
Surfaces rarely share a common period, and the force per area preventing
lateral motion averages to zero as the contact size grows
\cite{hirano93,muser01prl}.
In our simulations, $k^{Ia}_T$
depends strongly on $d'/d$, $\sigma'/\sigma$, $l_{\rm min}$,
and the orientations of the solids, and the total lateral stiffness
is generally too small to distinguish on the scale of Fig. \ref{fig:stiffload}.
Indeed,
this is desirable in comparing to continuum theories for
normal contact since they assume zero friction and lateral stiffness.

Our results for a wide range of parameters can be summarized by noting
that Eq. \ref{kappa}
arises because the distributions
of pressure and contact sizes 
are independent of load.
This implies $k^{Ia}_{T} = \alpha A_c c_{44}/\sigma$ where $\alpha$ is
of order one if the contact has the same stiffness as the substrate.
Since compliances add,
the total interfacial compliance is $1/k^{It}_T = 1/k^{Ia}_T+1/k^{Is}_T$.
Both contributions diverge as $F_N$ and $A_c$ go to zero,
explaining why $1/k^{It}_T$ can be
comparable to the bulk compliance of the bounding solids \cite{bowden86,bellow70}.
Since both contributions
scale with area, $k^{It}_T \propto A_c$
but with a lower slope than in Fig. \ref{fig:stiffarea}.

The substrate compliance dominates for sufficiently large systems
and roughness since $k^{Ia}_T/k^{Is}_T =
\alpha c_{44} h_{\rm rms} / \sigma E' \sqrt{| \nabla h|^2}  
 \sim (l_{\rm max}/l_{\rm min})^{H}$.
However, for small $\alpha$, the value of $l_{\rm max}$ required
to reach this limit may be large, particularly on the
scale of atomistic simulations, microelectromechanical systems,
or the wavelength of ultrasound used to measure shear stiffness \cite{experiments}.
For the case of $H=0.5$ in Fig. \ref{fig:stiffarea}, $k^I_{Ta}$ 
still reduces
the total interfacial stiffness by a factor of two
for $l_{\rm max} \sim 1300 \sigma \sim 400$nm with $\alpha=0.1$ and for
$l_{\rm max}  \sim 40\mu$m for $\alpha=0.01$.
Our measured $\alpha$ span this range and
one can estimate $\alpha$ in experimental systems from the
static friction coefficient $\mu_s$.
If $\beta \sigma$ is the lateral displacement for the force to reach
the static friction, then the above equations yield
$\alpha = \mu_s  \sqrt{| \nabla h|^2} /\kappa \beta$.
Typical values of $\beta$ are of order 1/4 so $\alpha$ and
$\mu$ are of the same order.

In conclusion, atomic scale simulations were used to study contact
between surfaces with roughness on a wide range of scales.
The results for area and normal stiffness are consistent with Persson's
continuum theory down to relatively small scales even though the solid
is not continuous or perfectly elastic.
The area and internal stiffnesses of systems with a range of $H$, $L$ and
$\nu$ show the linear scaling predicted in
Eqs. \ref{kappa} - \ref{eq:area}
with nearly constant values of $\kappa$ and $\gamma$.
The internal stiffnesses were shown to depend only on the
geometry of the contacting region.
Atomic scale displacements between contacting atoms
have little effect on the normal stiffness, but can
change the lateral stiffness by orders of magnitude.
This sensitivity makes lateral stiffness a promising
probe of the atomic scale interactions that underly friction.

Conversations with Martin M\"user and Bo Persson
are gratefully acknowledged.
This material is based upon work supported by
the Air Force Office of Scientific
Research under Grant No.~FA9550-0910232.


\begin{thebibliography}{29}
\expandafter\ifx\csname natexlab\endcsname\relax\def\natexlab#1{#1}\fi
\expandafter\ifx\csname bibnamefont\endcsname\relax
  \def\bibnamefont#1{#1}\fi
\expandafter\ifx\csname bibfnamefont\endcsname\relax
  \def\bibfnamefont#1{#1}\fi
\expandafter\ifx\csname citenamefont\endcsname\relax
  \def\citenamefont#1{#1}\fi
\expandafter\ifx\csname url\endcsname\relax
  \def\url#1{\texttt{#1}}\fi
\expandafter\ifx\csname urlprefix\endcsname\relax\def\urlprefix{URL }\fi
\providecommand{\bibinfo}[2]{#2}
\providecommand{\eprint}[2][]{\url{#2}}

\bibitem[{\citenamefont{Bowden and Tabor}(1986)}]{bowden86}
\bibinfo{author}{\bibfnamefont{F.~P.} \bibnamefont{Bowden}} \bibnamefont{and}
  \bibinfo{author}{\bibfnamefont{D.}~\bibnamefont{Tabor}},
  \emph{\bibinfo{title}{The Friction and Lubrication of Solids}}
  (\bibinfo{publisher}{Clarendon Press}, \bibinfo{address}{Oxford},
  \bibinfo{year}{1986}).

\bibitem[{\citenamefont{Greenwood and Williamson}(1966)}]{greenwood66}
\bibinfo{author}{\bibfnamefont{J.~A.} \bibnamefont{Greenwood}}
  \bibnamefont{and} \bibinfo{author}{\bibfnamefont{J.~B.~P.}
  \bibnamefont{Williamson}}, \bibinfo{journal}{Proc. R. Soc. London, Ser. A}
  \textbf{\bibinfo{volume}{295}}, \bibinfo{pages}{300} (\bibinfo{year}{1966}).

\bibitem[{\citenamefont{Persson}(2001)}]{persson01}
\bibinfo{author}{\bibfnamefont{B.~N.~J.} \bibnamefont{Persson}},
  \bibinfo{journal}{Phys. Rev. Lett.} \textbf{\bibinfo{volume}{87}},
  \bibinfo{pages}{116101} (\bibinfo{year}{2001}).

\bibitem[{\citenamefont{Hyun et~al.}(2004)\citenamefont{Hyun, Pei, Molinari,
  and Robbins}}]{hyun04}
\bibinfo{author}{\bibfnamefont{S.}~\bibnamefont{Hyun}},
  \bibinfo{author}{\bibfnamefont{L.}~\bibnamefont{Pei}},
  \bibinfo{author}{\bibfnamefont{J.-F.} \bibnamefont{Molinari}},
  \bibnamefont{and} \bibinfo{author}{\bibfnamefont{M.~O.}
  \bibnamefont{Robbins}}, \bibinfo{journal}{Phys. Rev. E}
  \textbf{\bibinfo{volume}{70}}, \bibinfo{pages}{026117}
  (\bibinfo{year}{2004}).

\bibitem[{\citenamefont{Pei et~al.}(2005)\citenamefont{Pei, Hyun, Molinari, and
  Robbins}}]{pei05}
\bibinfo{author}{\bibfnamefont{L.}~\bibnamefont{Pei}},
  \bibinfo{author}{\bibfnamefont{S.}~\bibnamefont{Hyun}},
  \bibinfo{author}{\bibfnamefont{J.-F.} \bibnamefont{Molinari}},
  \bibnamefont{and} \bibinfo{author}{\bibfnamefont{M.~O.}
  \bibnamefont{Robbins}}, \bibinfo{journal}{J. Mech. Phys. Sol.}
  \textbf{\bibinfo{volume}{53}}, \bibinfo{pages}{2385} (\bibinfo{year}{2005}).

\bibitem[{\citenamefont{Campa{\~n}\'a and M\"user}(2007)}]{campana07epl}
\bibinfo{author}{\bibfnamefont{C.}~\bibnamefont{Campa{\~n}\'a}}
  \bibnamefont{and} \bibinfo{author}{\bibfnamefont{M.~H.}
  \bibnamefont{M\"user}}, \bibinfo{journal}{Europhys. Lett.}
  \textbf{\bibinfo{volume}{77}}, \bibinfo{pages}{38005} (\bibinfo{year}{2007}).

\bibitem[{\citenamefont{Bellow and Nelson}(1970)}]{bellow70}
\bibinfo{author}{\bibfnamefont{D.~G.} \bibnamefont{Bellow}} \bibnamefont{and}
  \bibinfo{author}{\bibfnamefont{D.~D.} \bibnamefont{Nelson}},
  \bibinfo{journal}{Exp. Mech.} \textbf{\bibinfo{volume}{10}},
  \bibinfo{pages}{506} (\bibinfo{year}{1970}).

\bibitem[{\citenamefont{Persson}(2007)}]{persson07}
\bibinfo{author}{\bibfnamefont{B.~N.~J.} \bibnamefont{Persson}},
  \bibinfo{journal}{Phys. Rev. Lett.} \textbf{\bibinfo{volume}{99}},
  \bibinfo{pages}{125502} (\bibinfo{year}{2007}).

\bibitem[{\citenamefont{Persson}(2008)}]{persson08c}
\bibinfo{author}{\bibfnamefont{B.~N.~J.} \bibnamefont{Persson}},
  \bibinfo{journal}{J. Phys.: Condens. Matter} \textbf{\bibinfo{volume}{20}},
  \bibinfo{pages}{312001} (\bibinfo{year}{2008}).

\bibitem[{\citenamefont{Nagy}(1992)}]{experiments}
\bibinfo{author}{\bibfnamefont{M.} \bibnamefont{Gonzalez-Valadez}},
\bibinfo{author}{\bibfnamefont{A.} \bibnamefont{Baltazar}}, \bibnamefont{and}
  \bibinfo{author}{\bibfnamefont{R.~S.} \bibnamefont{Dwyer-Joyce}},
  \bibinfo{journal}{Wear} \textbf{\bibinfo{volume}{268}},
  \bibinfo{pages}{373} (\bibinfo{year}{2010}).


\bibitem[{\citenamefont{Archard}(1957)}]{archard57}
\bibinfo{author}{\bibfnamefont{J.~F.} \bibnamefont{Archard}},
  \bibinfo{journal}{Proc. R. Soc. London A} \textbf{\bibinfo{volume}{243}},
  \bibinfo{pages}{190} (\bibinfo{year}{1957}).

\bibitem[{\citenamefont{Persson and Ballone}(2000)}]{persson00}
\bibinfo{author}{\bibfnamefont{B.~N.~J.} \bibnamefont{Persson}}
  \bibnamefont{and} \bibinfo{author}{\bibfnamefont{P.}~\bibnamefont{Ballone}},
  \bibinfo{journal}{J. Chem. Phys.} \textbf{\bibinfo{volume}{112}},
  \bibinfo{pages}{9524} (\bibinfo{year}{2000}).

\bibitem[{\citenamefont{M\"user}(2008)}]{muser08prl}
\bibinfo{author}{\bibfnamefont{M.~H.} \bibnamefont{M\"user}},
  \bibinfo{journal}{Phys. Rev. Lett.} \textbf{\bibinfo{volume}{100}},
  \bibinfo{pages}{055504} (\bibinfo{year}{2008}).

\bibitem[{\citenamefont{Carbone and Bottiglione}(2008)}]{carbone08}
\bibinfo{author}{\bibfnamefont{G.}~\bibnamefont{Carbone}} \bibnamefont{and}
  \bibinfo{author}{\bibfnamefont{F.}~\bibnamefont{Bottiglione}},
  \bibinfo{journal}{J. Mech. Phys. Sol.} \textbf{\bibinfo{volume}{56}},
  \bibinfo{pages}{2555} (\bibinfo{year}{2008}).

\bibitem[{\citenamefont{Hyun and Robbins}(2007)}]{hyun07}
\bibinfo{author}{\bibfnamefont{S.}~\bibnamefont{Hyun}} \bibnamefont{and}
  \bibinfo{author}{\bibfnamefont{M.~O.} \bibnamefont{Robbins}},
  \bibinfo{journal}{Tribol. Int.} \textbf{\bibinfo{volume}{40}},
  \bibinfo{pages}{1413} (\bibinfo{year}{2007}).

\bibitem[{\citenamefont{Campa{\~n}\'a et~al.}(2008)\citenamefont{Campa{\~n}\'a,
  M\"user, and Robbins}}]{campana08}
\bibinfo{author}{\bibfnamefont{C.}~\bibnamefont{Campa{\~n}\'a}},
  \bibinfo{author}{\bibfnamefont{M.~H.} \bibnamefont{M\"user}},
  \bibnamefont{and} \bibinfo{author}{\bibfnamefont{M.~O.}
  \bibnamefont{Robbins}}, \bibinfo{journal}{J. Phys.: Condens. Matter}
  \textbf{\bibinfo{volume}{20}}, \bibinfo{pages}{354013}
  (\bibinfo{year}{2008}).

\bibitem[{\citenamefont{Benz et~al.}(2005)\citenamefont{Benz, Rosenberg,
  Kramer, and Israelachvili}}]{benz06}
\bibinfo{author}{\bibfnamefont{M.}~\bibnamefont{Benz}},
  \bibinfo{author}{\bibfnamefont{K.~J.} \bibnamefont{Rosenberg}},
  \bibinfo{author}{\bibfnamefont{E.~J.} \bibnamefont{Kramer}},
  \bibnamefont{and} \bibinfo{author}{\bibfnamefont{J.~N.}
  \bibnamefont{Israelachvili}}, \bibinfo{journal}{J. Phys. Chem. B}
  \textbf{\bibinfo{volume}{110}}, \bibinfo{pages}{11884}
  (\bibinfo{year}{2005}).

\bibitem[{\citenamefont{Lorenz and Persson}(2009)}]{lorenz09}
\bibinfo{author}{\bibfnamefont{B.}~\bibnamefont{Lorenz}} \bibnamefont{and}
  \bibinfo{author}{\bibfnamefont{B.~N.~J.} \bibnamefont{Persson}},
  \textbf{\bibinfo{volume}{21}}, \bibinfo{pages}{015003}
  (\bibinfo{year}{2009}).

\bibitem[{\citenamefont{Yang and Persson}(2008)}]{yang08}
\bibinfo{author}{\bibfnamefont{C.}~\bibnamefont{Yang}} \bibnamefont{and}
  \bibinfo{author}{\bibfnamefont{B.~N.~J.} \bibnamefont{Persson}},
  \bibinfo{journal}{Phys. Rev. Lett.} \textbf{\bibinfo{volume}{100}},
  \bibinfo{pages}{024303} (\bibinfo{year}{2008}).

\bibitem[{\citenamefont{Mindlin}(1949)}]{mindlin49}
\bibinfo{author}{\bibfnamefont{R.~D.} \bibnamefont{Mindlin}},
  \bibinfo{journal}{J. App. Mech.} \textbf{\bibinfo{volume}{16}},
  \bibinfo{pages}{259} (\bibinfo{year}{1949}).

\bibitem[{\citenamefont{Luan and Robbins}(2009)}]{luan09}
\bibinfo{author}{\bibfnamefont{B.}~\bibnamefont{Luan}} \bibnamefont{and}
  \bibinfo{author}{\bibfnamefont{M.~O.} \bibnamefont{Robbins}},
  \bibinfo{journal}{Tribol. Lett.} \textbf{\bibinfo{volume}{36}},
  \bibinfo{pages}{1} (\bibinfo{year}{2009}).

\bibitem[{\citenamefont{Luan and Robbins}(2005)}]{luan05}
\bibinfo{author}{\bibfnamefont{B.}~\bibnamefont{Luan}} \bibnamefont{and}
  \bibinfo{author}{\bibfnamefont{M.~O.} \bibnamefont{Robbins}},
  \bibinfo{journal}{Nature} \textbf{\bibinfo{volume}{435}},
  \bibinfo{pages}{929} (\bibinfo{year}{2005}),
  \bibinfo{journal}{Phys. Rev. E} \textbf{\bibinfo{volume}{74}},
  \bibinfo{pages}{026111} (\bibinfo{year}{2006}).

\bibitem[{\citenamefont{Hirano and Shinjo}(1993)}]{hirano93}
\bibinfo{author}{\bibfnamefont{M.}~\bibnamefont{Hirano}} \bibnamefont{and}
  \bibinfo{author}{\bibfnamefont{K.}~\bibnamefont{Shinjo}},
  \bibinfo{journal}{Wear} \textbf{\bibinfo{volume}{168}}, \bibinfo{pages}{121}
  (\bibinfo{year}{1993}).

\bibitem[{\citenamefont{M\"user et~al.}(2001)\citenamefont{M\"user, Wenning,
  and Robbins}}]{muser01prl}
\bibinfo{author}{\bibfnamefont{M.~H.} \bibnamefont{M\"user}},
  \bibinfo{author}{\bibfnamefont{L.}~\bibnamefont{Wenning}}, \bibnamefont{and}
  \bibinfo{author}{\bibfnamefont{M.~O.} \bibnamefont{Robbins}},
  \bibinfo{journal}{Phys. Rev. Lett.} \textbf{\bibinfo{volume}{86}},
  \bibinfo{pages}{1295} (\bibinfo{year}{2001}).

\bibitem[{lam()}]{lammps}
\bibinfo{note}{Simulations performed with LAMMPS http://www.cs.sandia.gov/~sjplimp/lammps.html}.

\bibitem[{\citenamefont{Campa{\~n}\'a and M\"user}(2006)}]{campana06}
\bibinfo{author}{\bibfnamefont{C.}~\bibnamefont{Campa{\~n}\'a}}
  \bibnamefont{and} \bibinfo{author}{\bibfnamefont{M.~H.}
  \bibnamefont{M\"user}}, \bibinfo{journal}{Phys. Rev. B}
  \textbf{\bibinfo{volume}{74}}, \bibinfo{pages}{075420}
  (\bibinfo{year}{2006}).

\bibitem{perssonpriv}
\bibinfo{author}{\bibfnamefont{C.}~\bibnamefont{Campa{\~n}\'a}},
  \bibinfo{author}{\bibfnamefont{B.~N.~J.} \bibnamefont{Persson}},
  \bibnamefont{and} \bibinfo{author}{\bibfnamefont{M.~H.}
  \bibnamefont{M\"user}}, \bibinfo{journal}{private communication.}

\end{thebibliography}
\end{document}